\newcommand{\jgr}{J. Geophys. Res.}
\newcommand{\grl}{Geophys. Res. Lett.}

\documentclass[11pt,twoside]{article}
\usepackage{asp2014}

\aspSuppressVolSlug
\resetcounters

\begin{document}

\title{ngVLA Observations of the Solar Wind}
\author{T. S. Bastian}
\affil{National Radio Astronomy Observatory, Charlottesville, VA, USA \email{tbastian@nrao.edu}}

% This section is for ADS Processing.  There must be one line per author.
\paperauthor{T. S. Bastian}{tbastian@nrao.edu}{0000-0002-0713-0604}{National Radio Astronomy Observatory}{}{Charlottesville}{VA}{22903}{USA}

%Please include a brief abstract that will be used by ADS for searching purposes.  
\begin{abstract}
The ngVLA has the potential to play a significant role in characterizing properties of the outer corona and the heating and acceleration of the solar wind into the inner heliosphere. In particular, using distant background sources to transilluminate the foreground corona and solar wind, a variety of radio propagation phenomena can be used to map plasma properties as a function of solar elongation and position angle throughout the solar cycle. These include angular broadening, interplanetary scintillations, and differential Faraday rotation, which can be used to map the solar wind velocity, determine properties of solar wind turbulence, and constrain the solar wind magnetic field. These observations will provide a global characterization of the solar wind that will be highly complementary to {\sl in situ} observations made by various spacecraft. In addition, such observations can be used to probe disturbances in the solar wind -- coronal mass ejections, for example -- that may impact the near-Earth environment.
\end{abstract}

\section{Introduction}

Although the idea of mass loss from the Sun dates back more than a century it was Parker (1958) who predicted the existence of a supersonic "solar wind" outflow from the Sun, subsequently confirmed in the early 1960s by spacecraft measurements.  To date, most of what we know about the solar wind is based on {\sl in situ} measurements of particles and fields from various vantage points in the solar system: from Earth orbit (e.g., the IMP satellites), missions near the L1 Langrangian point (e.g., ISEE-3, WIND, and ACE), those sent into the inner heliosphere (Helios 1 and 2),  those sent out of the ecliptic ({\sl Ulysses}), and those sent to the outer reaches of the heliosphere and into the local interstellar medium (Voyager 1 and 2) .  A new chapter of the exploration of the solar wind opens in the coming decade with the launch of the NASA {\sl Parker Solar Probe} (PSP), which will measure solar wind parameters at radial distances as close as  9.8 ${\rm R_\odot}$ from the Sun, or 0.045 AU. The PSP is complemented by the ESA {\sl Solar Orbiter} (SO), which will orbit the Sun at 60 ${\rm R_\odot}$ (0.28 AU) at an inclination of up to 34$^\circ$. 

Remote sensing observations are an important complement to {\sl in situ} measurements in the solar wind. Particularly important are white light coronagraphs (SOHO, STEREO/SECCHI) to trace solar wind outflows and disturbances therein (e.g., coronal mass ejections - CMEs), and UV coronagraph spectrometers (e.g., SOHO/UVCS) which have placed key constraints on electron, proton, and ion temperatures in the solar corona. Another powerful remote sensing technique is observations of radio propagation phenomena from the ground; that is, to observe a distant background radio source, spacecraft beacon, or other coherent radio source to measure the effects of the foreground solar wind plasma on the signal. These phenomena can be used to provide global context to {\sl in situ} measurements and can, moreover, measure solar wind properties in regions that are otherwise inaccessible to {\sl in situ} and other remote sensing techniques.  While radio propagation phenomena have been exploited since the earliest days of coronal  (Erickson 1964) and solar wind exploration (Hollweg \& Harrington 1968) they have generally been limited to experiments performed when a spacecraft was at an opportune location (superior conjunction), or when a bright background source (e.g., Tau A) was at a favorable solar elongation. Even Venus, illuminated by a radar signal, has been used as a background source (Coles \& Harmon 1989). These experiments have demonstrated the power and utility of radio propagation phenomena as probes of the solar wind and have placed valuable constraints on its properties.  However, these techniques have not been systematically exploited. Satellite beacons and radar signals are rarely available as probes and, until recently, radio instruments on the ground have not been sensitive enough to observe many background sources. With the advent of the ngVLA, the sensitivity barrier to the use of background sources is overcome. The sensitivity of the ngVLA will allow a large number of naturally occurring background sources to be used to systematically explore the heliosphere as discussed further in \S4.

In this chapter, the potential of the ngVLA to exploit several scattering phenomena as probes of the outer solar corona and the solar wind is outlined. The subject is vast and so the focus will be on just three propagation phenomena: angular broadening, interplanetary scintillations, and Faraday rotation. 

\section{Motivation}

Remarkable progress has been made in understanding the nature of the solar wind over the past fifty years (see, e.g., the monograph by Bruno \& Carbone 2016) but mysteries remain. It is widely believed that the solar wind undergoes spatially extended heating and that it plays a significant role in solar wind acceleration. Several heating mechanisms have been considered in detail including magnetic reconnection, wave dissipation, and turbulence. Turbulence has attracted significant attention in recent years as a means of not only heating the solar wind, but also mediating energy transport within the wind and modulating the transport of energetic particles. Briefly, the Sun is the source of outward propagating long-period Alfv\'en waves (AW). Sunward propagating AW can result from non-WKB wave reflection, velocity shear instabilities, and/or parametric instabilities which, in turn, interact with outward propagating AW to produce an (anisotropic) energy cascade to smaller spatial scales; i.e., a turbulent spectrum. When the cascade reaches scales of order the proton gyroradius the waves become compressive (kinetic Alfv\'en waves - KAW) and the wave energy is dissipated, thereby heating the ambient plasma (see, for example, Hollweg 1999; Dmitruk et al. 2002; Chandran et al. 2009; Salem et al. 2012). A detailed understanding of these processes nevertheless remains elusive and the fundamental questions remain germane:

\begin{itemize} 
\item{how are the corona and solar wind heated and how is the solar wind accelerated?}
\item{how does strong MHD turbulence develop and evolve in the solar wind?}
\item{what are the turbulence dissipation mechanisms?}
\item{what role does solar turbulence play in mediating energy deposition and transport? }
\item{how does solar wind turbulence modulate the transport of energetic particles (both solar energetic particles and cosmic rays)?}
\end{itemize}

\noindent These questions are relevant to other stars, of course. More broadly, the loss of mass and angular momentum due to outflows affect the evolution of a given star, and the extension of the wind from a star can profoundly influence the interstellar medium in its neighborhood. In order to address questions posed by the solar wind and their implications for other stars, our understanding of the solar wind must be placed on a comprehensive observational footing.  With the launch of PSP and SO, the inner heliosphere is a new frontier. For the first time, measurements will be made of the solar wind and solar wind turbulence close to the source. Nevertheless, there will still be many locations - the high latitude solar wind, for example - that will not be accessible to either spacecraft. Placing PSP/SO measurements into a broader context will be critically important. Particularly interesting is the region out to the Alfv\'en radius ($\sim 10-20$ R$_\odot$) where the wind becomes super-Alfv\'enic which, until the PSP mission, has been inaccessible to {\sl in situ} measurements. The nature of the solar wind, solar wind turbulence, and solar wind heating on either side of the Alfv\'en surface remains somewhat speculative at this time and awaits detailed characterization by both {\sl in situ} and remote sensing  observations. 

The ngVLA will be positioned to make significant and complementary contributions to this new frontier.  In particular, the ngVLA can leverage a number of radio propagation phenomena to provide comprehensive and complementary diagnostic observations of the solar wind in the inner heliosphere. Electron density fluctuations act as a ``passive scalar'' that traces solar wind turbulence through the spatial spectrum of electron density fluctuation, $\Phi_n(\kappa)$. The spectrum of density fluctuation cause variations in the plasma refractive index over a broad range of spatial scales. For radio waves propagating through a turbulent plasma medium, the fluctuations in the refractive index result in a variety of phenomena collectively referred to as ``scattering'' phenomena. We discuss three such phenomena here that the ngVLA will be able to effectively exploit. 

%\begin{table}[ [ht]
%\caption{Radio Propagation Phenomena}
%\smallskip
%\begin{center}
%\begin{tabular}{lc} 
%\tableline
%\noalign{\smallskip}
%Phenomenon & Plasma Properties Deduced\\
%\noalign{\smallskip}
%\tableline
%\noalign{\smallskip}
%Group delay & $\langle n_e\rangle \\
%Refraction & $\nabla n_e$ \\
%Angular broadening & $\Phi_{n_e}(q)$, PoS B orientation \\
%Intensity scintillations & $v_{sw}$, $\delta v_{sw}$ \\
%Faraday rotation & $B_\para$, $n_e$ \\
%Faraday fluctuations & $\delta B$
%\noalign{\smallskip}
%\tableline
%\end{tabular}
%}
%\end{center}
%\end{table}

\section{Angular Broadening}

Interferometric observations of angular broadening are particularly straightforward to perform and to interpret. The relevant observable is the mutual coherence function, $\Gamma(s)=\langle E({\bf r})E^\ast({\bf r+s})\rangle/\langle|E|^2\rangle$, where $E({\bf r})$ is the electric field measured by an antenna at location ${\bf r}$, and ${\bf s}$ is the projected baseline to another antenna located at ${\bf r+s}$. For a point-like radio source, $\Gamma(s)$ is related to the visibility function $V(s)$ and the wave structure function, $D_\circ(s)=\langle[\phi(\bf{r})-\phi(\bf{r+s})]^2\rangle$, as $\Gamma(s) = V(s)/V(0) = \exp{[-D_\circ(s)/2]}$, which is related to the angular spectrum of the source through a Fourier transform (Fig. 1, left panel). A loss of phase coherence with increasing interferometer baseline, $s$, due to plasma turbulence leads to greater values of $D(s)$ which manifests itself as angular broadening of the source in the image plane. A convenient form for the spatial spectrum of turbulence (Coles et al. 1987) is a power law with an inner scale at $s = l_\circ$, given by $\Phi_n(\kappa)=C_n^2\kappa^{-\alpha-2}\exp[-(\kappa l_\circ/2)^2]$, where $C_n^2$ is the turbulence level (Rickett 1977). In this case, $D_\circ(s)$ has particularly simple asymptotic limits: when $s > l_\circ$, $D_\circ(s) \propto C_n^2\lambda^2 s^\alpha \Delta z$, and for $s < l_\circ$ we have $D_\circ(s)\propto C_n^2\lambda^2l_\circ^{\alpha-2} s^2 \Delta z$ where $\lambda$ is the wavelength of the radio radiation and $\Delta z$ is the thickness of the turbulent slab. The angular size of the scattered source is given by $\theta_s\sim \lambda/r_{diff}$, where $s=r_{diff}$ is the coherence scale (or diffractive scale) defined by $D_\circ(r_{diff}) = 1$. 

\articlefiguretwo{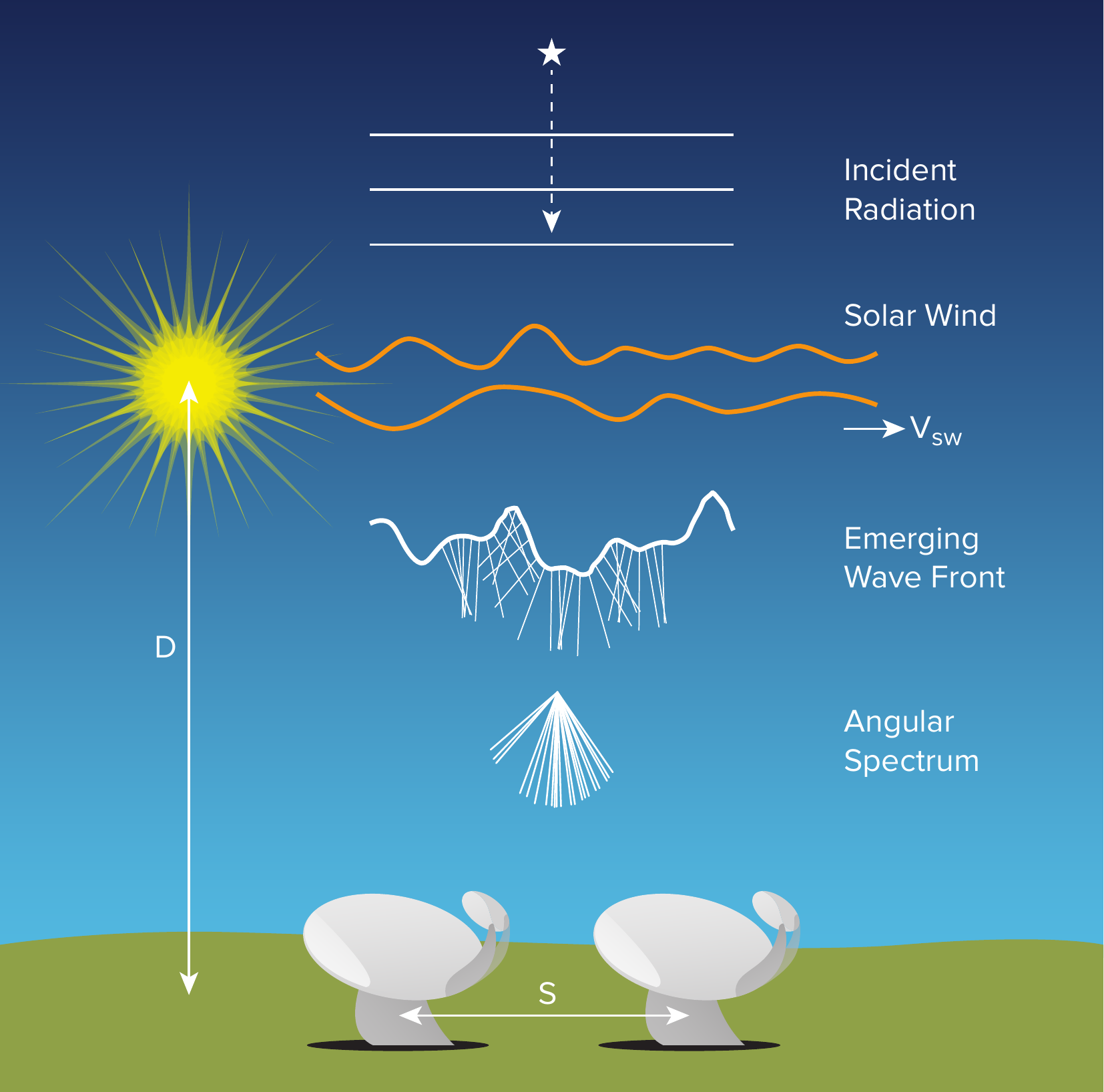}{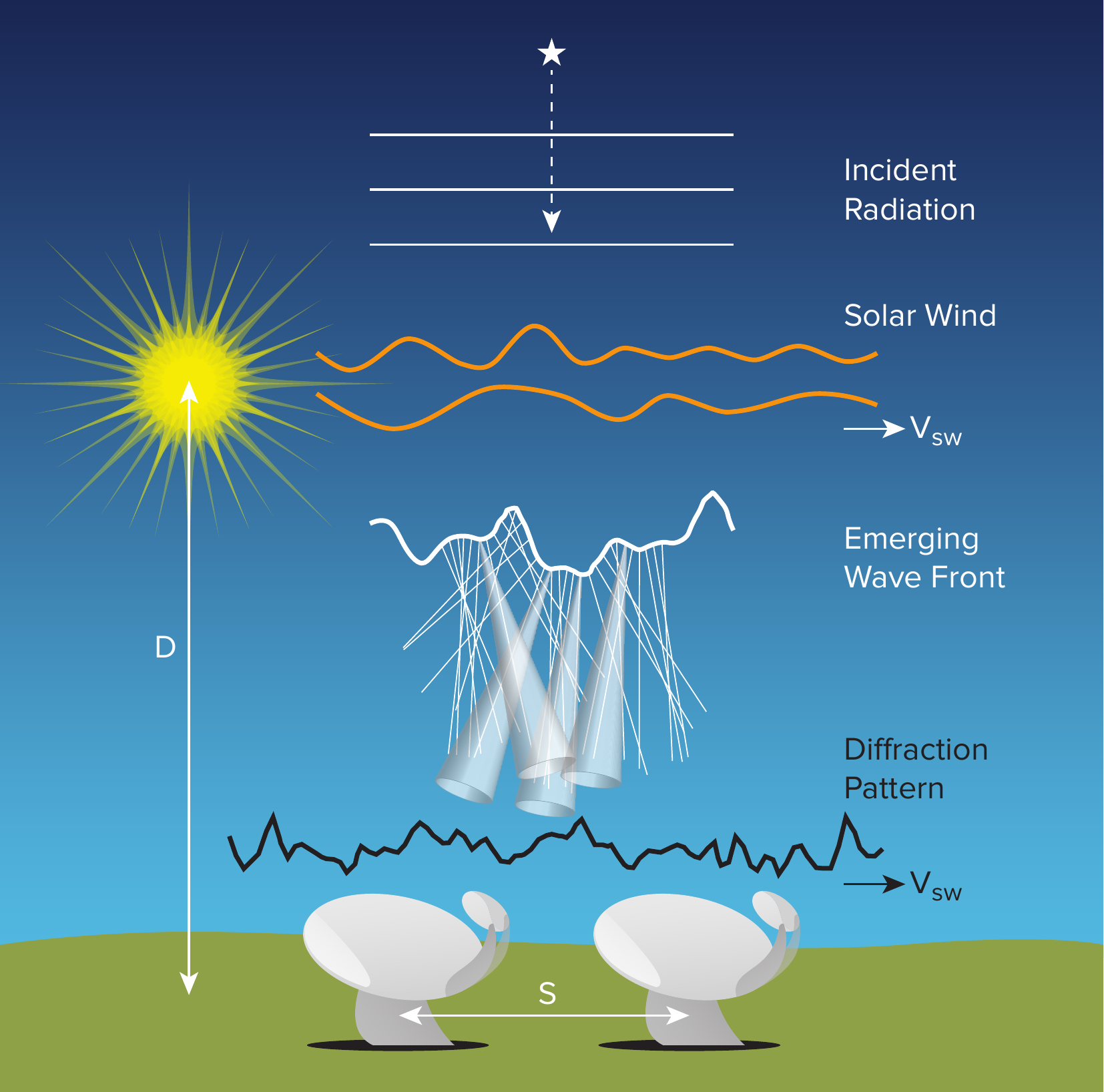}{fig:swab}{Left: A cartoon illustrating the interaction of radio waves from a distant point source interacting with foreground solar wind turbulence. The incident plane waves emerge as a corrugated wave front and propagate $D=1$ AU to the observer. The wave normals represent different angles of arrival of the radiation observed on the ground as described by the angular spectrum. Right: the solar wind effectively scatters radiation into a cone of angular width $\lambda/r_{diff}$ that illuminates a patch on the ground of size $\sim D \lambda/r_{diff}$. These patches constructively and destructively interfere to create a diffraction pattern that sweeps over the observer with the solar wind velocity $V_{sw}$.}

Angular broadening measurements of a background source or, equivalently, measurements of the structure function $D_\circ(s)$ can therefore be used to deduce properties of solar wind turbulence through $\Phi_{n}(\kappa)$ for the range of transverse scales $s$ measured by an interferometer. When many background sources are observed over a range of solar elongations and position angles, a map of the solar wind properties in the inner heliosphere can be constructed. An advantage of angular broadening measurements is that they can be made under conditions of strong or weak scattering (see below). 

Such work was undertaken by the VLA in the 1990s (e.g., Armstrong et al. 1990; Anantharamaiah et al. 1994; Bastian 1999) using small numbers of VLA calibrator sources to transilluminate the solar wind. These observations show that the solar wind turbulence in the inner heliosphere is highly anisotropic, with turbulent eddies drawn out along the magnetic field. The degree of anisotropy may decrease at elongations $>\!6$ R$_\odot$ and may also decrease on spatial scales $>\!100$ km (Grall et al. 1997) but this has not been systematically studied. Intriguingly, there are indications that the degree of anisotropy may be smaller on short spatial scales than on longer spatial scales, perhaps indicative of an isotropic dissipation mechanism (Armstrong et al. 1990). Since the density inhomogeneities are anisotropic along the magnetic field, the scattering is enhanced perpendicular to the direction of the magnetic field. Angular broadening observations can therefore be used to infer the orientation of the coronal and solar wind magnetic field in the plane of the sky. Furthermore, while well described by a power-law on spatial scales corresponding to VLA baselines in the A configuration (kilometers to 10s of kilometers), VLA observations showed that the power-law index $\alpha$ on these scales is significantly smaller than the Kolmogorov value of 5/3 observed on much larger spatial scales using {\sl in situ} spacecraft measurements and VLBI phase scintillation measurements (Coles \& Harmon 1989; Coles et al 1991). The scale on which the spectrum flattens is of order 100-300 km in the inner solar wind. The flat part of the spectrum represents a significant excess of turbulent power relative to that expected for an extrapolation of the Kolmogorov spectrum observed on large spatial scales (1000s of km to $10^5$ km), consistent with the onset of KAW turbulence on small spatial scales (Chandran et al. 2009). 

VLA observations were strongly constrained by its limited sensitivity to a handful of strong sources. The JVLA has much greater sensitivity, allowing a significant increase in the number of sources that may be exploited for "solar wind tomography" (Fig. 2). However, the JVLA remains limited in its baseline coverage to spatial scales of just a few 10s of km. The ngVLA will provide both enormous sensitivity and baseline coverage out to $\sim\!1000$ km which will allow not only angular broadening measurements to be fully exploited (see \S6), but also enables additional propagation phenomena to be more fully exploited of which two will now be briefly discussed. 

\articlefigure{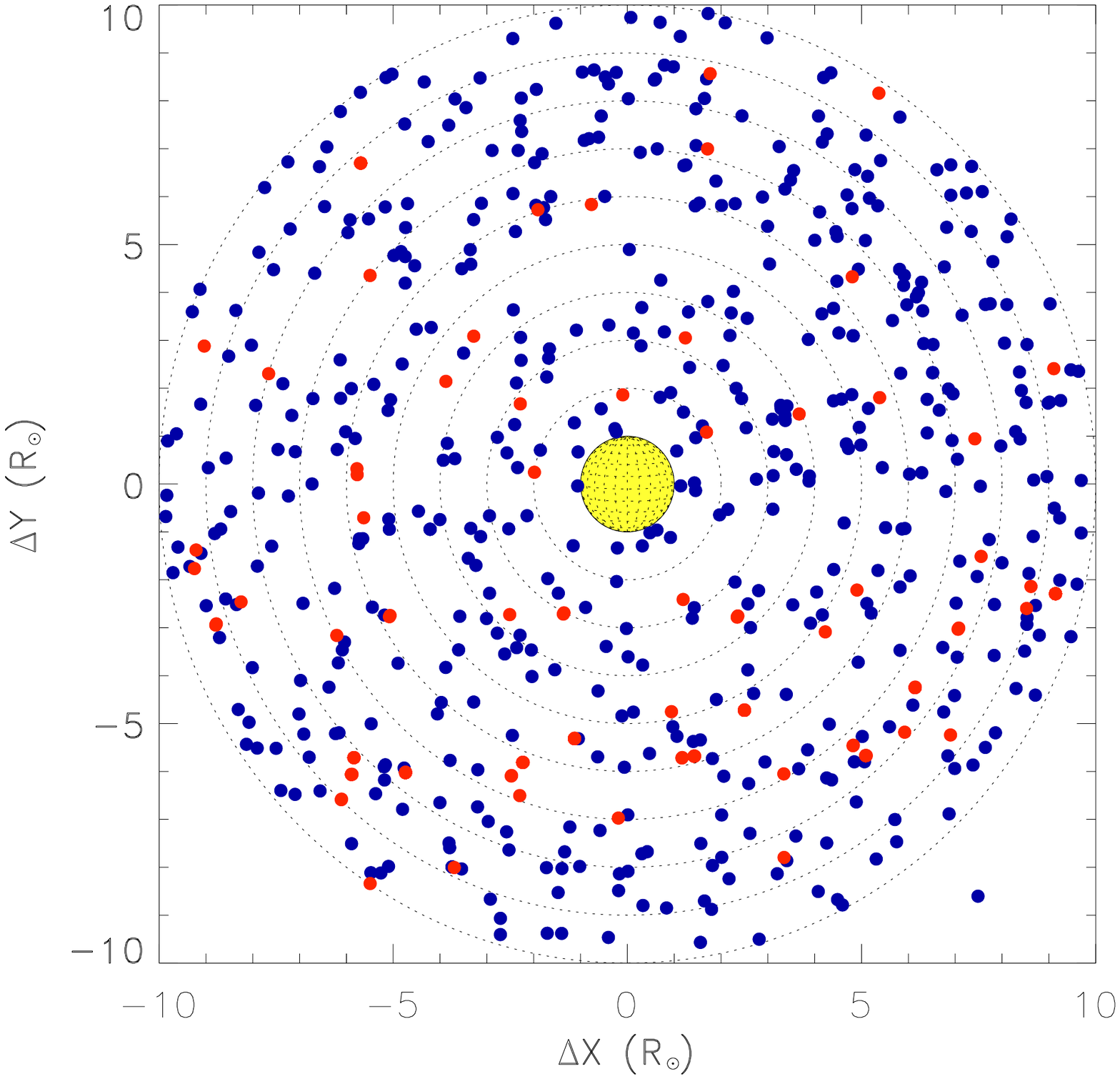}{fig:sources}{An illustrative plot of background sources that could be used as probes of the Sun's corona and solar wind. The blue points show daily positions (local noon) of {\sl Jodrell Bank - VLA Astrometric Survey} 8.4 GHz sources (Patnaik et al. 1992; Browne et al. 1998; Wilkinson et al. 1998; Winn et al. 2003) that pass within 10 R$_\odot$ of the Sun during the course of one year. These are compact, flat-spectrum sources dominated by milliarcsec cores that have flux densities $>50$ mJy. The red points show all known millisecond pulsars (MSPs) that likewise pass within 10 R$_\odot$ of the Sun (Manchester et al. 2005). Note that the bulk of the MSPs are near the galactic plane and are therefore accessible during a relatively short window of time each year. }

\section{Interplanetary Scintillations}

A second scattering phenomenon is interplanetary scintillation (IPS), intensity variations as a function of time. It is useful to distinguish between two regimes: weak scattering and strong scattering, demarcated by comparing $r_{diff}$ to the Fresnel scale $r_F=\sqrt{\lambda D}$, where $D=1$ AU. In weak scattering, $r_{diff}\gg r_F$ and intensity fluctuations are caused by weak focusing and defocusing of density fluctuations on the Fresnel scale and $\sigma_I/\langle[\rangle \ll1$. The time scale of IPS in the weak scattering regime is therefore of order $r_F/V_{sw}\sim$ few $\times 0.1$s where $V_{sw}$ is the solar wind speed in the plane of the sky. In strong scattering $r_{diff}\ll r_F$.  Inhomogeneities with a scale $r_{diff}$ scatter radiation into a diffractive cone of angular width $\lambda/r_{diff}$ which, at 1~AU illuminates a patch on the ground of size $L=\lambda D/r_{diff}$. A given point on the ground is therefore illuminated by many phase-independent patches which interfere with one another, producing a diffraction pattern on the ground (Fig. 1, right panel). The solar wind motion carries the diffraction pattern over an observer at the solar wind speed, causing intensity variations on a time scale $r_{diff}/V_{sw}\sim$ few $\times$ 0.01s.  

Two observers separated by a distance $s$, oriented parallel to the outflow, can independently measure the time series of scintillations and cross correlate them to deduce the solar wind speed at a given solar elongation and position angle. A range of baselines and orientations can be used to deduce the solar wind velocity, including non-radial velocities, projected onto the plane of the sky. Observations of many background sources can map out the solar wind velocity and acceleration in the inner heliosphere. The solar wind is in fact comprised of fast solar wind streams ($\sim\!700$ km/s) that emerge from magnetically open regions in the solar corona (coronal holes), and slow solar wind ($\sim\!400$ km/s). Hence, two different velocities may be present along a given line of sight. Klinglesmith (1997), Breen et al. 2006, and others have shown that IPS measurements made on sufficiently well-separated baselines ($\sim\!500-1000$ km) can be used to resolve and measure the speeds of both types of solar wind. In addition, since the solar wind carries Alfv\'en wave fluctuations these, too, can be constrained (Klinglesmith 1997). IPS is most useful as a diagnostic in the weak scattering regime, where the IPS amplitude $\sigma_I/\langle I\rangle \propto n^2 \lambda^2\propto \lambda^2/r^4$; in strong scattering, it saturates with $\sigma_I/\langle I\rangle\approx 1$. Observations in the weak scattering regime can be assured by selecting appropriate values for $\lambda$ and/or $r$. 

A second type of scintillation, referred to as ``refractive'' scintillation, occurs in the strong scattering regime. It results from weak focusing and defocusing effects on the scale of the scatter-broadened source at 1 AU: $r_{ref} = D\theta_s=\lambda D/r_{diff}=r_F^2/r_{diff}$.  Refractive scintillation occurs on a time scale $r_{ref}/V_{sw}\sim\!10$ s. In the strong scattering regime we have $r_{diff} < r_F < r_{ref}$. Hence, refractive scintillation provides a second means of leveraging a scattering phenomenon to probe solar wind properties -- in particular,  those of $\Phi_n(\kappa)$ -- on scales of $\sim\!1000s$ km.  Specifically, Goodman \& Narayan (1989) show that refractive scintillation yields a power law tail in mean square visibility $\langle |V(s)|^2\rangle$ on baselines $s>r_{diff}$ in the so-called  ``average'' visibility regime for shallow spectra ($\alpha<2$). This can be measured as part of any angular broadening observing program, allowing simultaneous constraints to be placed on $\Phi_n(\kappa)$ on scales of order $r_{diff}$ and $r_{ref}$ as long as the source is observed in the strong scattering regime (Bastian 1999).

\section{Faraday Rotation}

Faraday rotation (FR) is the phenomenon where the plane of polarization of a linearly polarized source -- a spacecraft beacon or a naturally occurring background source -- rotates as it traverses a magnetized plasma. The rotation $\Delta\chi$  is in the amount

\begin{equation}
\Delta\chi=\lambda^2\ \Biggl[\biggl( {{e^3}\over{2\pi m_e^2 c^4}} \biggr)\ \int n(z) {\bf B}(z)\cdot dz\Biggr] =\lambda^2\ {\rm RM}
\end{equation}

\noindent where the term in brackets is referred to as the {\sl rotation measure} (RM). FR can be used to constrain both the coronal and solar wind magnetic field and density. The most straightforward experiment is to measure the gross properties of the magnetic field by making a differential measurement: the intrinsic position angle of the polarization vector compared with that after passage through the coronal plasma or the solar wind. P\"atzold et al. (1987) observed the linearly polarized carrier signal of the Helios spacecraft during occultations for nearly a decade, allowing them to constrain the magnetic field and its time variation between 3-10 $R_\odot$. Ingleby et al. (2007) used the VLA to observe a number extragalactic radio sources with known polarization properties through the solar wind plasma at elongations ranging from approximately 5-10 $R_\odot$. In order to extract magnetic field constraints, however, independent observations of the solar wind density and/or careful modeling of the solar wind density are required. A second, but similar use of differential FR measurements is to observe the passage of a CME across a polarized background source, again using spacecraft beacons (Bird et al. 1985) or ground-based radio telescopes (Jensen \& Russell 2008, Kooi et al. 2017). Ancillary observations or models are needed here, too, to disentangle the number density from the magnetic field. 

Another intriguing use of FR measurements is to constrain the spectrum of Alfv\'en waves in the solar wind turbulence, believed to play a role in heating and momentum deposition. Two approaches have been attempted: i) observations of temporal FR fluctuations (FRF); and, ii) FR depolarization as a result of turbulence introducing spatial FRF across the beam of a radio telescope. The first of these approaches was exploited by Efimov et al. (1993) and Andreev et al. (1997) who used FRF measurements of the Helios-2 beacon to measure FRF spectra, which they interpreted in terms of long-period Alfv\'en waves. Mancuso \& Spangler (1999) used the VLA to search for FRF as evidence for long-period Alfv\'en wave, finding a marginal result of lower than expected magnitude. Spangler \& Mancuso (2000) also used this approach to search for short-period Alfv\'en waves. However, no evidence for depolarization due to FRF was found although the sensitivity of the experiment was marginal. 

Finally, the scientific potential of millisecond pulsars (MSPs) as a probe of the solar wind is strong. The advantage of using pulsars is that they provide both a polarized signal with which to measure the RM and also a precision pulse from which the group delay or, equivalently, the {\sl dispersion measure} (DM) can be deduced:

\begin{equation}
\tau_G=\lambda^2\ \Biggl[\biggl( {{e^2}\over{2\pi m_e c^3}} \biggr)\ \int n(z) dz \Biggr] = \lambda^2\ {\rm DM}
\end{equation}

\noindent  Together, the RM and DM can be used to impose strong constraints on the solar wind density and magnetic field. You et al. (2012) used the Parkes Pulsar Timing Array to observe several MSPs and were able to make simultaneous RM and DM measurements at elongations ranging from 6-10 R$_\odot$. 

\subsection{ngVLA Observations}

The key capability offered by the ngVLA is global mapping of critical solar wind properties in two dimensions over the course of the solar activity cycle.  This will provide critical benchmarks against which models of solar wind heating, acceleration, and turbulence must be measured. Here, necessary ngVLA capabilities are briefly summarized and the key science goals are reiterated under the assumption that ngVLA performance will be closely aligned with the reference instrument described by Selina et al. (2018). 

The key capabilities that will allow the ngVLA to make unique contributions to solar wind studies are its wide and continuous frequency coverage, its sensitivity, and its fixed baseline coverage. Past experience has shown that it is unlikely that frequencies much larger than 20 GHz will be needed for solar wind studies. Hence, the emphasis will be on ngVLA reference bands 1-3. The sensitivity of the ngVLA will be critical for mapping a variety of solar wind parameters in two dimensions. Such mapping depends on having adequate numbers of point-like background sources (or sources for which there are excellent models) to probe the foreground corona and solar wind. Fig. 2 displays suitable sources that are currently accessible by the JVLA based on the JVAS survey of compact, flat spectrum sources $>\!50$ mJy (blue points) at 8.4 GHz during the course of one year. The sensitivity of the JVLA is such that a 10 min observation (Goodman \& Narayan's ``ensemble average'' regime) yields a typical rms of 10-20 $\mu$Jy in bands up to 20~GHz, more than sufficient to observe JVAS sources with high SNR. The ngVLA will be 5-10 times more sensitive than the JVLA at a given frequency and similar bandwidth and will therefore be able to use background sources of 10~mJy or lower as effective probes although these will need to be surveyed prior to use. 
%Source count surveys suggest that at any given time there will be $\sim\!1000$ background sources within 20 R$_\odot$ of the Sun at any given time, of which a significant fraction will be unresolved by the ngVLA (e.g., roughly a quarter of background sources at 1.5 GHz will be unresolved; J. Condon, private comm.)

Also important is the greatly expanded baseline coverage provided by the ngVLA. The reference ngVLA provides 214 antenna with spacings up to $\sim\!1000$ km, nearly a factor of 30 longer than is currently possible with the JVLA. These will enable measurements of the spatial spectrum $\Phi_n(\kappa)$ over a much larger range of scales, including the inflection point in the spectrum at spatial scales of $\sim\!100$ km.  The ngVLA will also provide unparalleled opportunity for IPS measurements. The reference instrument places 46 ngVLA antennas on baselines greater than the current footprint of the JVLA providing more than 500 baselines sampling scales $>\!200$ km.  Hence, a long baselines can be found that are parallel to the solar wind vector for a source at essentially any elongation and position angle. It is important to note that the fixed antenna configuration allows observations at any time of the year. They will no longer be constrained by the VLA configuration cycle, allowing a continuous record of solar wind properties to be measured. While the reference antenna configuration includes baselines out to 1000 km the distribution is weighted more heavily toward the EW quadrants -- the number of baselines exceeding 500 km in the NS quadrants is relatively small and further optimization of the array configuration therefore may be needed. 

To address solar wind science objectives, the following ngVLA measurements are key:
\begin{itemize}
\item {\sl Solar wind turbulence}: The angular size of scatter-broadened background sources depends on wavelength and elongation as $\sim \lambda^2/R^2$ where $R$ is measured in units of R$_\odot$, and the angular resolution of the array varies as $\theta\propto \lambda$. Scaling from results reported by Anantharamaiah et al (1994) at $\lambda=6$~cm for R=1.7-16 R$_\odot$, one finds that the ngVLA will be able to probe angular broadening out to elongations R$\sim 60$ R$_\odot$. Hence, unlike the VLA/JVLA, the ngVLA will allow observations over a greater range of elongations and over a much greater range of $s$ (and, hence, $\kappa$), providing measurements of the turbulence level, the evolution of $\Phi_n(\kappa)$ from a Kolmogorov-like spectrum to a flatter-than-Kolmogorov spectrum, its degree of anisotropy, the inner scale $l_{in}$ - all as functions of solar elongation and position angle. Particularly important to wave-heating theories will be the use of the ngVLA as probe of solar wind properties as it transits the Alfv\'en radius, at $\sim\!10-20\ R_\odot$, where the solar wind speed becomes super-Alfv\'enic . Models based on Alfv\'enic turbulence (Cranmer \& van Ballegooijen 2005) suggest that both velocity and density fluctuations reach a maximum near the Alfv\'en surface.  IPS observations through the transsonic region (e.g., Lotova 1988) also suggest enhanced scattering there. 
\item {\sl Solar wind velocity and acceleration}: IPS observations offer the means of mapping solar wind velocities throughout the inner heliosphere. Unlike previous measurements for which point measurements were occasionally made of solar wind speed, the ngVLA will enable two dimension mapping observations of the solar wind speed and acceleration in the inner heliosphere for a given phase in the solar cycle, critical for understanding the relationship between coronal heating and solar wind heating and acceleration. For measurements of IPS, total power measurements will be needed at each antenna at a sampling rate of 100 Hz. It will be desirable to phase up proximate antennas and to cross-correlate antenna groups rather than individual antennas as a means of improving the sensitivity of the IPS measurements.
\item {\sl Solar wind magnetic field}: Since solar wind inhomogeneities are drawn out along the local magnetic field they are highly anisotropic. This results in anisotropic scattering which, in turn, enables allows observers to determine the mean line-of-sight orientation of the magnetic field in the plane of the sky toward the source in question. By observing many sources, a map of the the magnetic field orientation in the inner heliosphere can be constructed. When MSPs are available for observation, angular broadening measurement plus RM and DM measurements will be available to constrain both the magnetic field vector and magnitude in the inner heliosphere. For FR/FRF measurements, polarimetric measurements are needed over wide bandwidths, with the lower two bands favored (1.2-3.5 GHz and 3.5-12.3 GHz). The excellent sensitivity and the anticipated fidelity of polarimetric imaging will allow far deeper measurements to be made of FRF and Faraday depolarization.
\item {\sl Space weather}: All of the above phenomena may be used to probe space weather drivers such as co-rotating interaction regions (CIRs) and coronal mass ejections (CMEs). CIRs are due to long-lived fast solar wind stream that are experienced as recurrent structures at Earth as the Sun rotates with its Carrington period of 27.3 days. CMEs involve the eruption of a magnetic flux rope, driving up to $\sim\!10^{16}$ gm of material at speeds exceeding 2000 km/s into the interplanetary medium. Both CIRs and CMEs can interact with the Earth's magnetosphere, producing geomagnetic storms. Large CMEs drive shocks that accelerate solar energetic particles to high energies that are detrimental to both manned and unmanned spacecraft. Studies of CIRs and CMEs in the inner heliosphere will provide new insights into so-called ``research to operations'' aspects of space weather forecasting and ``now-casting''. 

% 5.2 x 10^4 /ster at 1.5 GHz --> 0.661 deg beam --> 10 / bm
% 3.2 at 3 GHz --> 0.165 deg beam --> 1.6
% 2.0 at 6 GHz --> 0.041 deg beam --> 0.25

\end{itemize}

With few exceptions, the reference ngVLA concept is already an excellent match to requirements for solar wind and space weather studies. Support of IPS observations require care in avoiding self-generated RFI signals and system time constants that could affect cross-correlations between antennas or subarrays defined by groups of antennas. Requirements for supporting these studies otherwise require careful planning, pilot studies, and surveys to supplement and characterize JVAS and other background sources, and to identify larger numbers of polarized sources for FR/FRF observations. While a given observation may not require significant telescope time, many observations of background sources will be needed to support comprehensive mapping of solar wind properties in the inner heliosphere. The most fruitful work will likely require systematic studies through large proposals.

\acknowledgments I thank Dustin Madison for providing a curated list of MSPs drawn from the {\sl ATNF Pulsar Catalogue} and both Dustin Madison and Scott Ransom for discussion about the use of pulsars for RM and DM measurements. 

%\bibliography{editor}  % For BibTex

% For non-BibTex:

\end{document}